\documentclass[10pt,onecolumn]{article}
\usepackage{graphicx}
\usepackage{amsmath}
\usepackage{amssymb}
\usepackage{siunitx} 
\usepackage{color} 
\usepackage[numbers,square,sort&compress]{natbib} 
\graphicspath{{Figures/png/}} 

\title{Backcoupling of acoustic streaming on the temperature field inside high-intensity discharge lamps}

\author{\normalsize Joerg Schwieger$^{*1,2}$,  Bernd Baumann$^{1}$,  Marcus Wolff$^{1}$, \\ \normalsize Freddy Manders$^{3}$, and Jos Suijker$^{3,4}$}

\date{\footnotesize
	$^{1}$Hamburg University of Applied Sciences, Department of Mechanical Engineering and\hfill \phantom{.}\\
	\hspace{0.4mm} Production, Berliner Tor 21, 20099 Hamburg, Germany\hfill \phantom{.}\\
	$^{2}$University of the West of Scotland, School of Engineering and Computing, High Street, \hfill \phantom{.}\\
	\hspace{1mm} Paisley PA1 2BE, United Kingdom \hfill\phantom{.}\\
	$^{3}$Philips Lighting, Steenweg op Gierle 417, 2300 Turnhout, Belgium \hfill\phantom{.}\\
	$^{4}$Technical University Eindhoven, Den Dolech 2, 5612AZ Eindhoven, Netherlands \hfill\phantom{.}\\
	$^*$ Joerg.Schwieger@HAW-Hamburg.de\hfill\phantom{.}}
\begin{document}
\maketitle

\begin{abstract}
Operating high-intensity discharge lamps in the high frequency range ($\num{20}-\SI{300}{kHz}$) provides energy-saving and cost reduction potentials. However, commercially available lamp drivers do not make use of this operating strategy because light intensity fluctuations and even lamp destruction are possible. The reason for the fluctuating discharge arc are acoustic resonances in this frequency range that are excited in the arc tube. The acoustic resonances in turn generate a fluid flow that is caused by the acoustic streaming effect. Here, we present a 3D multiphysics model to determine the influence of acoustic streaming on the temperature field in the vicinity of an acoustic eigenfrequency. In that case a transition from stable to instable behavior occurs. The model is able to predict when light flicker can be expected. The results are in very good accordance with accompanying experiments.
\end{abstract}

\section{Introduction}
\label{sec:10_Introduction}
Approximately one fifth of the world production of electric power is consumed by \num{30} to \num{33} billion electric light sources \cite{Zissis.2010,Stoffels.2006,Waide.2006}. High intensity discharge (HID) lamps represent a main part of the artificial light sources. They are used for street and shop lighting as well as automobile headlights and other applications. A superior color rendering index combined with a sun-like luminance characterizes HID lamps \cite{NEMA.2010}. These  advantages over light emitting diodes will guarantee a market for HID lamps in the future.
\begin{figure} 
	\centering
	\includegraphics[width=0.7\linewidth,trim=0 0 0 0,clip]{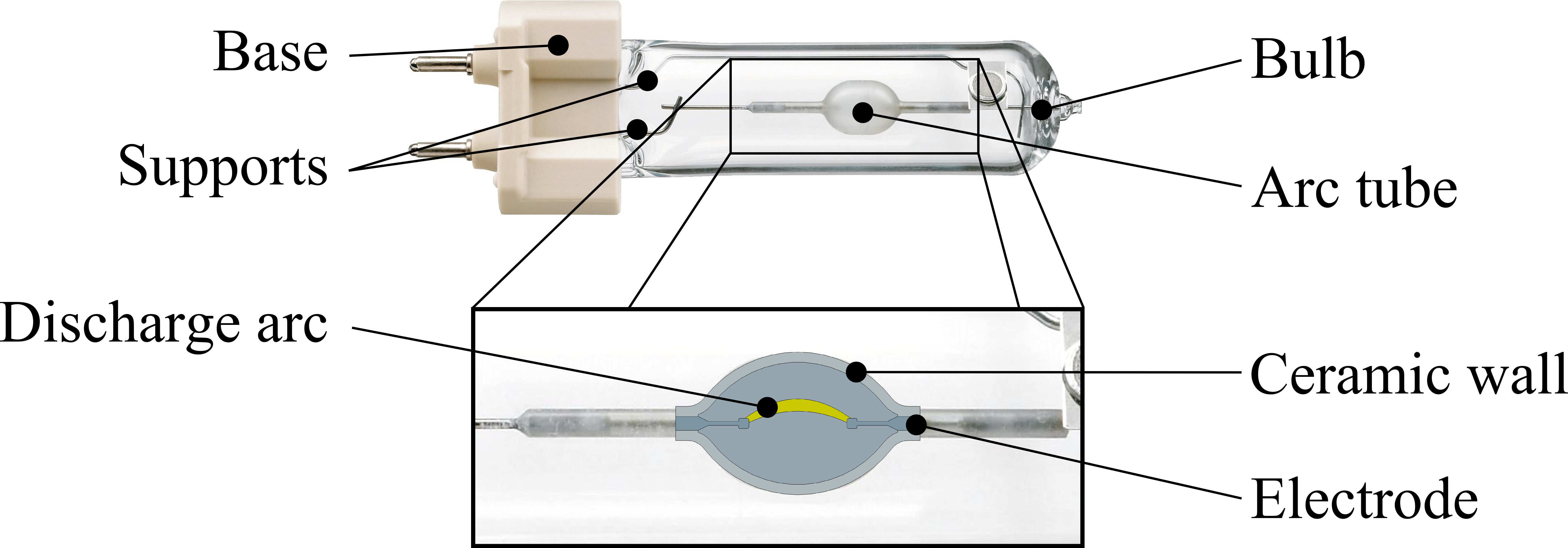}
	\caption{Setup of the investigated HID lamp (Philips \SI{35}{W} \num{930} Elite) in horizontal operation position.\label{fig:StraightBowedArc_TNR}}
\end{figure}
The structure of the low-wattage HID lamp, that is investigated in this article, is shown in figure~\ref{fig:StraightBowedArc_TNR}. The distance between the tungsten electrodes is \SI{4.8}{mm}. The tube from polycrystalline alumina (PCA) contains a mixture of argon, mercury and metal halides. A voltage is applied to the electrodes to establish a plasma arc of high temperature inside the arc tube. In order to minimize material costs of the electronic driver, the lamp should be operated in the high frequency range of \SIrange{20}{300}{kHz} \cite{Trestman.2002}. However, the periodic ohmic heating  excites acoustic resonances in this frequency range. The resonances can induce low frequency plasma arc fluctuations of approximately \SI{10}{Hz} \cite{Schwieger.2014b} that are visible as light flicker. The acoustic streaming (AS) phenomenon is the link between the high frequency acoustic resonance and the low frequency flicker \cite{Afshar.2008}.

To investigate the light flicker phenomenon, a three-dimensional stationary finite element model of this lamp was developed. Initially, it calculates the electric potential, the buoyancy driven velocity field and the temperature field. On the basis of the temperature field the acoustic eigenmodes are determined. Thereafter, the acoustic pressure field and the AS force are determined for the second mode. With the aid of this model, a symmetry breaking transition for the flow field was detected \cite{Baumann.2014b}. Here, we present results of the backcoupling of the superimposed AS and buoyancy force on the velocity and the temperature field.

\section{Model}
\label{sec:20_Model}
The finite element model for the numerical investigation of the horizontally operated HID lamp comprises the arc tube wall, its filling and both electrodes, and it solves several coupled partial differential equations. All simulations were executed with COMSOL Multiphysics and MATLAB.

\subsection{Geometry and mesh}
\label{sec:21_GeometryAndMesh}
\begin{figure} 
	\centering
	\includegraphics[width=0.6\linewidth,trim=0 0 0 0,clip]{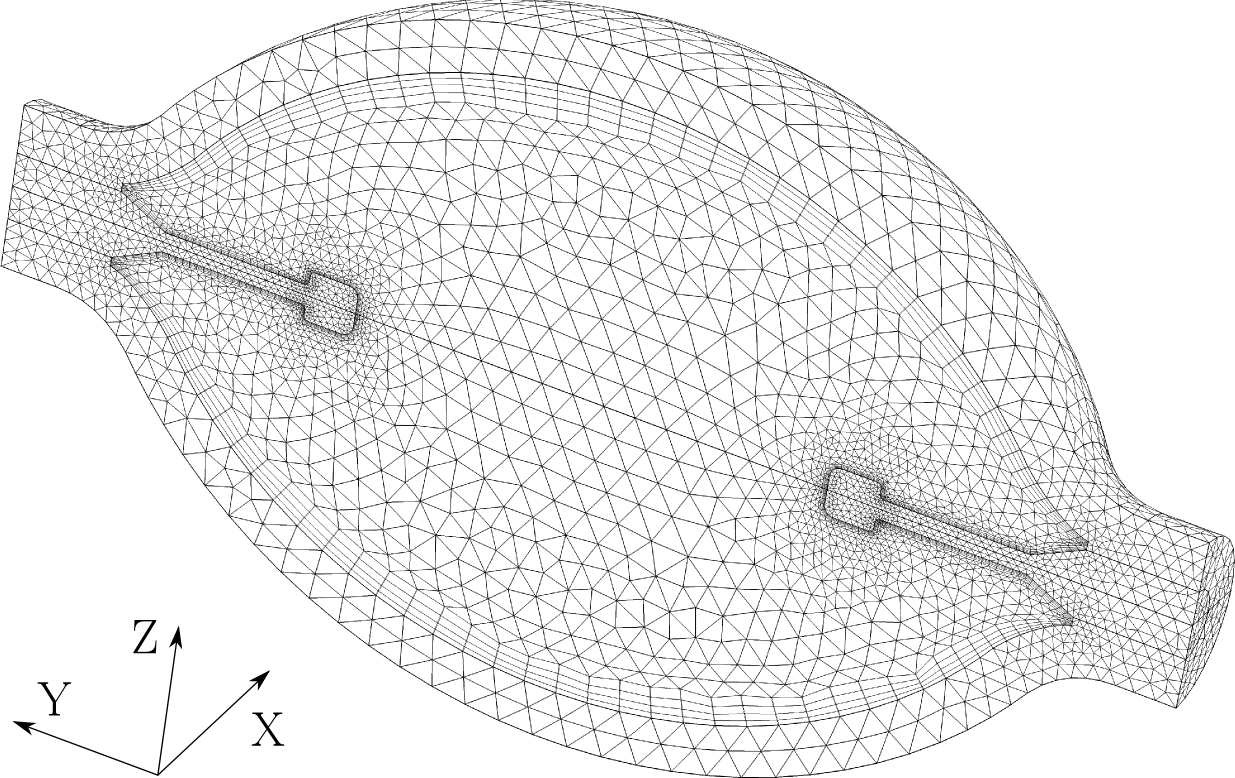}
	\caption{Finite element mesh of the half model. The origin of the coordinate system is located in the center between the electrodes.}
	\label{fig:FE-Mesh1}
\end{figure}

The model of the HID lamp was created with slight simplifications. The coils at the electrodes were simplified to cylinders to reduce the number of finite element nodes and, therefore, the computing time. The lamp is tilted by \SI{5}{\degree} against the horizontal axis to prevent stability problems due to instable paths related to the symmetry breaking transition that was observed in a prior investigation \cite{Baumann.2014b}.
The model has to be designed in three dimensions in order to consider the upward bending of the plasma arc due to buoyancy. The temperature field is symmetrical to the vertical $y$-$z$-plane (figure~\ref{fig:FE-Mesh1}). Therefore, only one half of the physical geometry has to be considered in this simulation step. The temperature field is mirrored at the $y$-$z$-plane to map it onto the full geometry and to calculate the acoustic modes.

The finite element mesh (figure~\ref{fig:FE-Mesh1}) used for the simulation of the half model consists of approximately \num{123000} elements. In the boundary layer of the arc tube prism elements were used, whereas the rest of the model is composed of tetrahedral elements.
The accuracy of the solution was tested with different mesh resolutions. The results obtained with the selected mesh differ from results with a mesh with \num{164000} elements by less than \SI{1}{\%}. The computing time needed with the mesh displayed in figure~\ref{fig:FE-Mesh1} could be reduced by a factor of \num{10} compared to the simulation with the very fine mesh.

\subsection{Temperature Field}
\label{sec:22_TemperatureField}
The coupled differential equations of mass, momentum, charge and energy conservation were used to obtain the temperature distribution. The differential equation of charge conservation
\begin{equation} \vec{\nabla}\cdot \left( \sigma\vec{E} \right) = 0 \label{eq:CurrentConservation}\end{equation}
enables the calculation of the electric field $\vec{E}$. The electric conductivity $\sigma$ and other material parameters are listed in a previous publication \cite{Baumann.2014b}. Because no essential differences are expected, a direct current (dc) was implemented instead of an alternating current (ac)\footnote{At ac condition the direction of the carrier wave periodically changes. On time average the system is symmetric. The dc boundary conditions have been set up accordingly.}. The inward current density on the electrodes was obtained by the electrode tip area and the electric current $I_{\rm lamp}$. With the nominal power of the lamp $P_{\rm lamp}$ and the voltage drop between the electrodes $U$ the electric current $I_{\rm lamp} = P_{\rm lamp}/U$ was computed. The tube's wall and the symmetry plane were modeled as electric insulators.

For Newtonian fluids the stationary velocity field $\vec{u}$ is described by the Navier-Stokes equation
\begin{equation} \rho \left( \vec{u}\cdot\vec{\nabla} \right) \vec{u}=\vec{\nabla}\cdot \left[-P\mathbf{I}+\eta \left( \vec{\nabla}\vec{u}+ \left( \vec{\nabla}\vec{u} \right) ^{\rm T} \right)-\frac{2}{3}\eta\left(\vec{\nabla}\cdot\vec{u}\right)\mathbf{I}\right] + \vec{f} \label{eqNavierStokes}\end{equation}
and mass conservation
\begin{equation} \vec{\nabla}\cdot(\rho\vec{u})=0 \label{eq:MassConservation}\end{equation}
($\mathbf{I}$ identity matrix, $\rho$ density, $P$ pressure inside the arc tube, $\eta$ dynamic viscosity). Without AS the velocity field is driven by the gravitational force only. Its force density is described by
\begin{equation} f_l=-\delta_{l3}\rho g \end{equation}
($\delta_{l3}$ Kronecker symbol: index \num{3} corresponds to the $Z$-direction, $g=\SI{9.81}{m/s^2} $). The boundary condition $\vec{u}=\vec{0}$ was used at the tube's wall and the electrodes. At the symmetry plane $\vec{n}\cdot\vec{u}=0$ ($\vec{n}$ normal vector) was applied and in the center the static pressure $P=\SI{2.81}{MPa}$ was specified\footnote{To preserve model symmetry, the center point was chosen. The position of the static pressure point is of no importance.}.

The Elenbaas-Heller equation
\begin{equation}\vec{\nabla}\cdot \left(-\kappa\vec{\nabla}T \right)+\rho c_p \vec{u}\cdot \vec{\nabla}T=\sigma |\vec{E}|^2-q_{\rm rad} \label{eq:ElenbaasHeller}\end{equation}
($\kappa$ thermal conductivity, $c_p$ specific heat capacity, $q_{\rm rad}$ density of power radiation) determines the temperature field $T$ inside the arc tube, of the wall and of the electrodes. On the external surface of the wall heat is emitted by radiation. This is described by the boundary condition
$\vec{n}\cdot(-\kappa\vec{\nabla}T)=\epsilon \sigma_{\rm SB} \left( T^4_{\rm amb}-T^4 \right)$ ($\epsilon$ emissivity, $\sigma_{\rm SB}$ Stefan-Boltzmann constant, $T_{\rm amb}$ ambient temperature). On the symmetry plane the emissivity is set to zero (no radiation). At the electrode grounds a constant temperature of $T=\SI{1450}{K}$ was defined. The acoustic pressure field in the arc tube can be calculated with the aid of the resulting temperature field and the power density of heat generation.

\subsection{Acoustic Streaming}
\label{sec:23_AcousticStreaming}
For the calculation of the AS force the acoustic pressure in the arc tube has to be calculated first. To obtain the acoustic pressure $p$, the Helmholtz equation
\begin{equation} \vec{\nabla} \cdot \left( \frac{1}{\rho}\vec{\nabla}p\right)+ \frac{\omega^2}{\rho c^2} p=i\omega \frac{\gamma-1}{\rho c^2} \mathcal{H}	\label{eq:InhomogenousHelmholtz} \end{equation}
has to be solved ($c$ speed of sound, $\gamma$ heat capacity ratio). The source term of the Elenbaas-Heller equation ${\cal H}=\sigma |\vec{E}|^2-q_{\rm rad}$ describes the power density of heat generation. The arc tube filling is treated as an ideal gas with the speed of sound field
\begin{equation} c = \sqrt{\kappa \frac{RT}{M}} . \end{equation}
The wall of the arc tube and the electrodes are treated as sound hard. The solution of the Helmholtz equation is expressed by an acoustic mode. It is described elsewhere, how the pressure amplitude, that takes the loss mechanisms of heat conduction and viscosity on the surface as well as in the volume into account, is calculated \cite{Kreuzer.1977,Baumann.2007,Baumann.2009}.

The AS force arises when a standing sound wave in the arc tube of the HID lamp is excited \cite{Afshar.2008}. In addition to an oscillatory motion, an averaged fluid motion is induced. The motion is a consequence of the viscous force that develops due to the shear viscosity of the fluid particles. It generates vortex-like motions that are distinguished by their location: Outer (or Rayleigh) streaming occurs outside the boundary layer and inner streaming occurs inside the boundary layer \cite{Boluriaan.2003}.

The AS force is implemented into the source term of the Navier-Stokes  equation. Instead of the gravitational force only, the force density is now
\begin{equation}\label{eq:ASforce} f_l = \frac{\partial \overline{\rho v_k v_l}}{\partial x_k} - \delta_{l3}\rho g \end{equation}
($\vec{v}$ sound particle velocity). In this equation Einstein's summation convention is used and the overbar indicates time averaging. In the case of time harmonic waves the force density simplifies to
\begin{equation} f_l = \frac{1}{2} \frac{\partial \rho \hat{v}_k \hat{v}_l}{\partial x_k} - \delta_{l3}\rho g. \end{equation}
 $\vec{\hat{v}}$ denotes the amplitude of the sound particle velocity. It can be obtained from the acoustic pressure \cite{Temkin.1981}:
\begin{equation} \vec{\hat{v}} \left( \vec{r}, \omega\right) = \frac{1}{{\rm i} \omega \rho} \vec{\nabla}p \left( \vec{r}, \omega \right)  .\end{equation}
The velocity drop in the boundary layer had to be considered additionally because the mode calculation does not incorporate this condition. Therefore, the sound particle velocity was multiplied with a factor that drops in the boundary layer from one to zero \cite{Schuster.1940,Hamilton.2003b}. The implementation of this function into the model has previously been described \cite{Baumann.2014b}.

\subsection{Simulation procedure}
\label{sec:24_SimulationProcedure}
The calculation of the temperature, the acoustic pressure and the AS force field have to be coupled in order to get the stationary temperature field at a certain frequency. Since the discharge arc is excited at this frequency, it is called \emph{excitation frequency}. 

Initially, the velocity and temperature fields as well as the acoustic mode and eigenfrequencies are calculated with the gravitational force only. On the basis of the acoustic modes the pressure amplitude is calculated. With the aid of the amplitude, the AS force at a specified excitation frequency is computed. 
In order to include the back coupling of the AS field on the other fields, a recursion loop was implemented. From the second iteration on, the fluid flow is driven by buoyancy and the AS force. This leads to a different velocity and temperature field compared to the first step. The acoustic eigenmode and the resonance frequency change and lead to an altered AS force that serves as input for the next step.
This recursion procedure is continued until the results of the current and the previous step coincide within a certain limit. The resonance frequency is used as a convergence criterion. Once the resonance frequency changes by less than \SI{5}{Hz}, the solution is considered converged.

The fields obtained in this way are used as initial condition for the next excitation frequency. Since we are mainly interested in results near a resonance, the excitation frequency is increased when the resonance frequency is higher and decreased when the resonance frequency is lower than the excitation frequency. The step size is adopted to the difference of both frequencies. The closer the resonance and the excitation frequency are, the higher the pressure amplitude and AS force. Because the AS force depends quadratically on the acoustic particle velocity (see Equation~\ref{eq:ASforce}), the force increases significantly near the resonance frequency. Hence, a very small step size of the excitation frequency has to be chosen.

\section{Experimental setup}
\label{sec:30_Experiments}
In the experiment the operating voltage of the lamp consists of two components. The square-wave part with a carrier frequency $f_c$ of \SI{400}{Hz}  is used to operate the lamp at stable conditions. Flicker of the discharge arc is excited by a high frequency sinusoidal part. This excitation frequency $f_e$ is varied in the range of \SIrange{41}{44}{kHz}. The combined voltage input is described by ($\omega_e=2\pi f_e$, $\omega_c=2\pi f_c$)
\begin{equation}U(t) = \begin{cases} A\left[+1+\alpha\sin \left( \omega_e t \right) \right] \hspace{1cm} 2n<\omega_ct\leq2n+1 \\ A\left[-1+\alpha\sin \left( \omega_e t \right) \right] \hspace{1cm} 2n+1<\omega_ct\leq2n+2 \end{cases} n\in\mathbb{N}_0.\end{equation}
The amplitude $A$ is adjusted to the nominal power of the lamp. The ratio of the square-wave to the sinusoidal-wave voltage amplitude is described by the modulation depth $\alpha$. In the experiments it is set to a constant value that is just high enough to stimulate arc flicker. A more detailed description of the experimental setup can be found elsewhere \cite{Schwieger.2014}.

Before the measurements the lamp is operated at stable conditions ($\alpha=0$) for at least \num{20} minutes to reach a stationary state. Then, the modulation depth is set to a constant value and, simultaneously, the excitation frequency is set to its initial frequency some kHz off the resonance frequency. This initial value is either \SI{41}{kHz} when the frequency is afterwards increased or \SI{44}{kHz} when the frequency is decreased. These operating conditions are kept constant for \SI{20}{s}. During this time the voltage drop between the electrodes is measured \num{40} times. At the initial step no acoustic resonance is excited so that the mean value of the measured voltages can be defined as the basic voltage. In the next step the excitation frequency is shifted by \SI{50}{Hz} into the direction of the resonance frequency. To prevent lamp failures, which are caused by too high arc tube temperatures or by temperature oscillations of the arc tube, the voltage measurements are from then on used to control the experiment. If one of the \num{40} measurement exceeds the basic voltage by \SI{8}{V} or if the voltage fluctuates by more than \SI{1.5}{V}, the experiment is terminated. If these criteria are not fulfilled, the experiment is continued until the excitation frequency reaches \SI{41}{kHz} and \SI{44}{kHz}, respectively.

\section{Results}
\label{sec:40_Results}
\subsection{Acoustic modes and eigenfrequencies}
\label{sec:41_AcousticEigenmodes}

Initially, the \SI{5}{\degree} tilted model without AS was solved. The simulation computes a lowest acoustic eigenfrequency of \SI{33100}{Hz} which is not apparent in the experiment. At \SI{47750}{Hz} and \SI{48130}{Hz} two neighboring azimuthal modes appear\footnote{If gravity is switched off, the two modes are degenerate.}. In this article the simulations, which include the backcoupling of AS on all other fields, focus on the mode at \SI{47750}{Hz}. The experiment shows the influence of both eigenfrequencies.

The excitation frequency was set to \SI{47750}{Hz} and the recursion started. Once convergence was reached, the eigenfrequency shifts to \SI{47400}{Hz} due to AS. As described above, the excitation frequency for the next step has to be decreased by \SI{50}{Hz}. The resulting eigenfrequency is \SI{40}{Hz} below the previous value at \SI{47360}{Hz}. Reducing the excitation frequency again by \SI{50}{Hz} further decreases the difference between excitation and eigenfrequency so that the simulation of the next step is conducted closer to the resonance. This results in a stronger AS force. Continuing in this way the simulations were successful down to an excitation frequency of \SI{46313}{Hz} where the difference to the resonance frequency diminishes to \SI{55}{Hz}. At even lower frequencies the simulation did not converge. We assume that no time-independent solution exists at these frequencies so that stationary simulations are inappropriate\footnote{A similar situation is known from the Taylor-Couette system, where the stationary Taylor cells undergo a Hopf bifurcation to the time-periodic wavy vortex state \cite{Cross.2009}.}. To test this assumption, we plan to perform a linear stability analysis in the near future.

\subsection{Voltage drop}
\label{sec:42_VoltageDrop}

A quantity, which is easy to determine in the simulation as well as in the experiment, is the voltage drop between the electrodes. In order to demonstrate the reliability of the FE model, the voltage of the simulations and experiments are compared. 

\begin{figure} 
	\centering
	\includegraphics[width=0.9\linewidth,trim=0 0 0 0,clip]{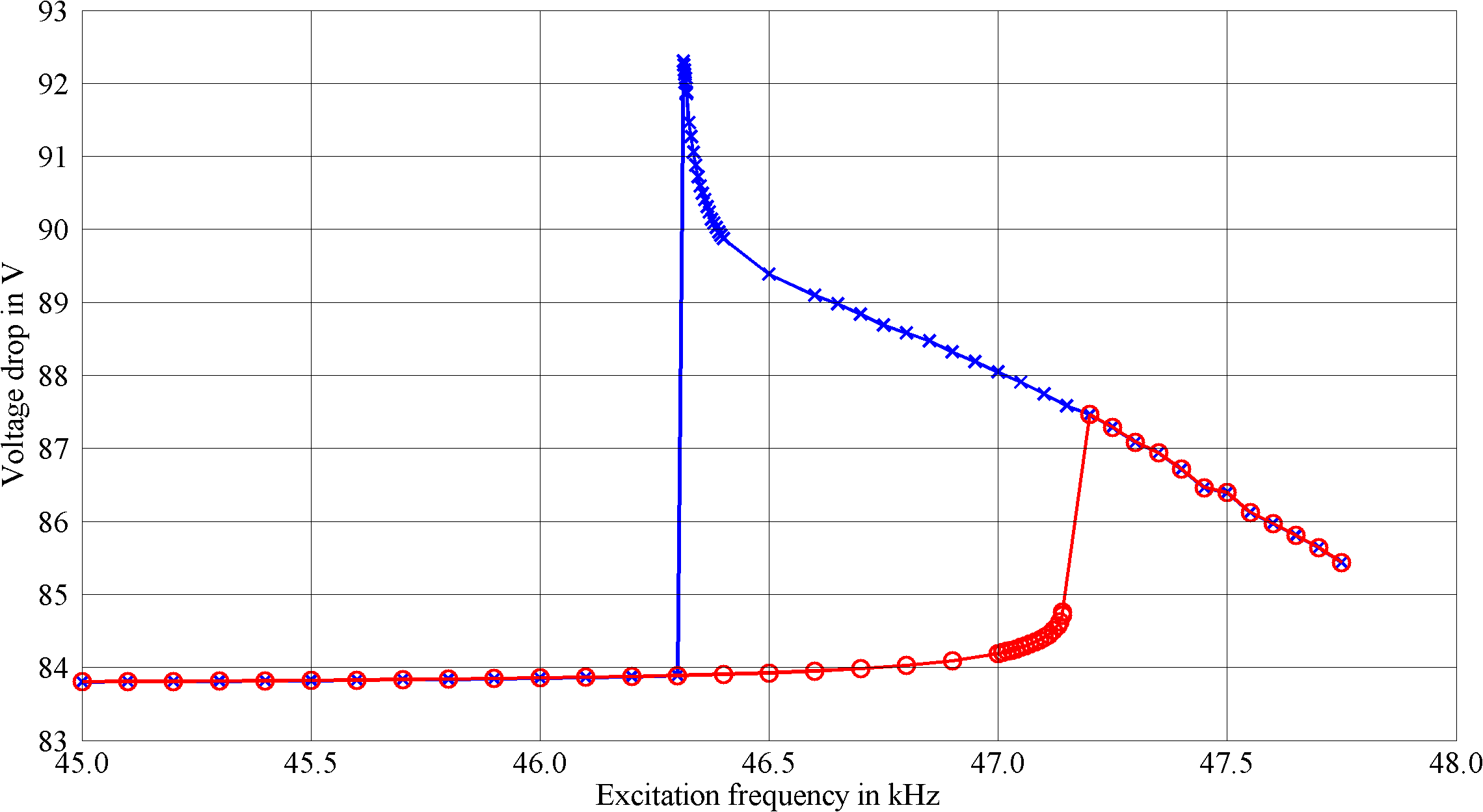}
	\caption{Simulation result of the voltage drop between the electrodes over the excitation frequency. The blue crosses mark voltage drops at decreasing excitation frequency. The red circles indicate voltage drops at increasing excitation frequencies.\label{fig:VoltvsXF}}
\end{figure}

Figure~\ref{fig:VoltvsXF} shows the simulation results of the voltage as function of the excitation frequency. The two curves correspond to the behavior of the voltage for increasing and decreasing excitation frequency, respectively. When the frequency is decreased from \SI{47.8}{kHz} to \SI{46.3}{kHz}, a significant voltage increase of \SI{7 }{V} occurs. A further lowering of the frequency leads to a voltage jump to values coinciding with the values of the curve representing the increasing frequency simulation. At increasing excitation frequency the voltage stays constant at about \SI{84}{V} for the frequency range from \SI{45.0}{kHz} to \SI{47.0 }{kHz}. Then the voltage increases by about \SI{1}{V}, before it jumps to the upper curve. The data clearly demonstrate the appearance of a hysteresis, which is expected in a nonlinear system.
The situation is similar to the one of a Duffing oscillator: The resonance frequency depends on the amplitude. If the oscillation is strong enough, the equation, which couples the amplitude and the excitation frequency, is cubic and its solution includes three real roots in a certain frequency range. The peak of the response curve either tilts to low or high frequencies and overlaps the response curve describing the amplitude far off the resonance \cite{Thompson.1986}.

\begin{figure} 
	\centering
	\includegraphics[width=0.9\linewidth,trim=0 0 0 0,clip]{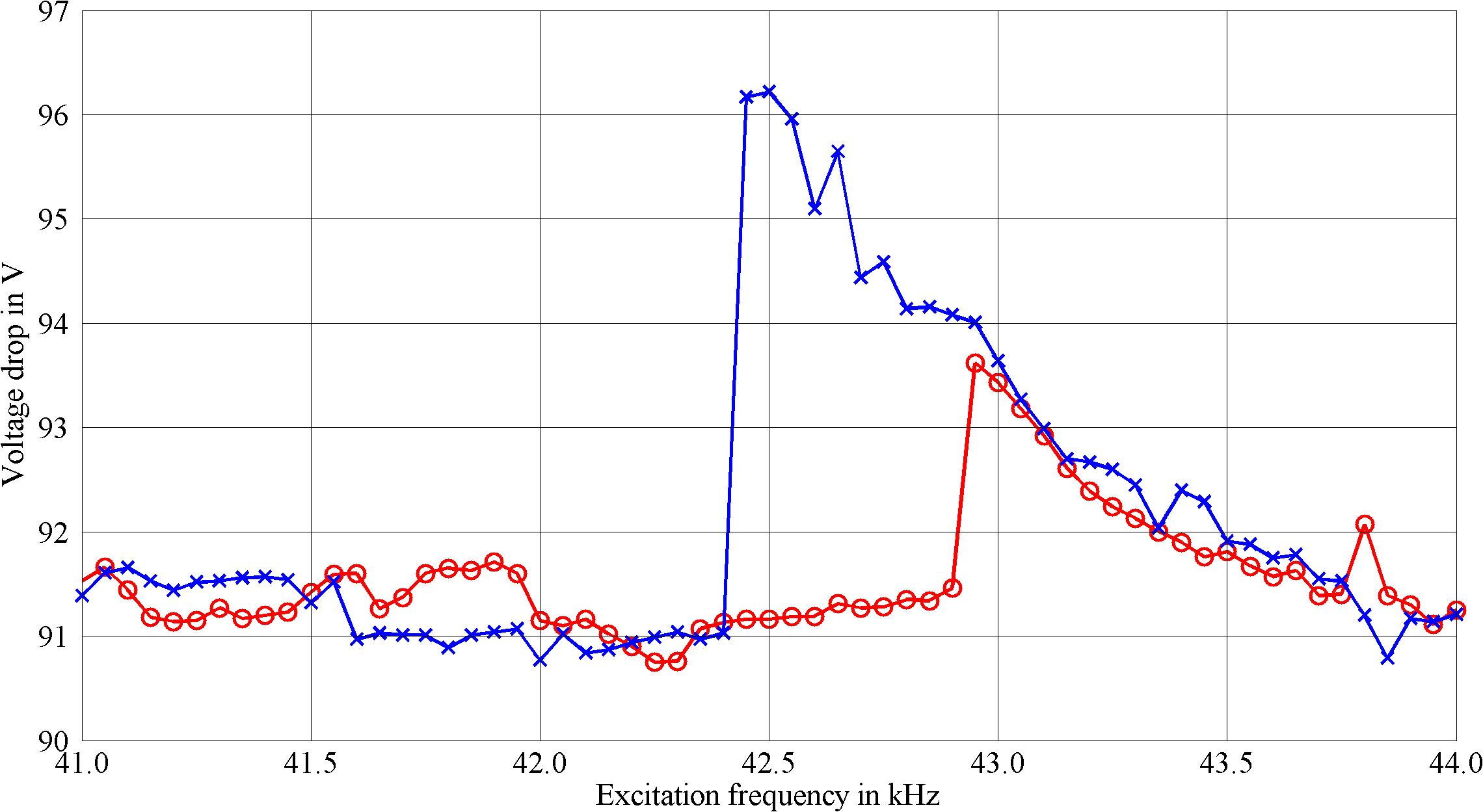}
	\caption{Experimental results of the voltage drop between the electrodes over the excitation frequency at a modulation depth of \SI{2}{\%}. Red circles indicate measurements at increasing and blue crosses at decreasing excitation frequency.\label{fig:VoltExp}}
\end{figure}

Figure~\ref{fig:VoltExp} shows the results of the experimental voltage measurements over the excitation frequency. The frequency deviation between the experimental and the simulation result of \SI{4}{kHz} is related to the simplified geometry and uncertainties in the material properties \cite{Schwieger.2014,Schwieger.2013}. Additionally, some boundary conditions, e.g. the constant temperature at the electrode ground or the uniformly distributed electric current density on the electrodes, only approximate the conditions in the lamp.
For a decreasing excitation frequency the voltage increases up to \SI{96}{V} at \SI{42.4}{kHz}. Here, the experimental results show a voltage jump like the one observed in the simulation. This jump has also been observed in high pressure sodium lamps \cite{Chhun.2011}. The significant voltage increase is related to the excitation of an acoustic mode. In contrast to the simulation, in which only one of the neighboring acoustic modes is considered, figure~\ref{fig:VoltExp} shows repercussions of both modes. The small voltage peak at the frequency of \SI{43.8}{kHz} indicates the second mode.

The simplified geometry and uncertainties in the material properties are also responsible for the different voltages: In the simulation the voltage far off an acoustic resonance is \SI{83.8}{V} and in the experiment approximately \SI{91.5}{V}. Apart from these differences in the absolute values the graphs in figure~\ref{fig:VoltvsXF} and figure~\ref{fig:VoltExp} show excellent agreement.

The voltage jump at an increasing excitation frequency is \SI{2.5}{V} compared to \SI{2.3}{V} in the experiment. At a decreasing excitation frequency the voltage in the experiment increases by \SI{5.1}{V} and in the simulation by \SI{8.3}{V}. The frequency range of the hysteresis (voltage peak at an increasing to the voltage peak at a decreasing excitation frequency) is \SI{840}{Hz} in the simulation and \SI{500}{Hz} in the experiment.
It would be interesting to measure the hysteresis at different modulation depths to determine its influence on the voltage jump and the jump frequencies. 


\subsection{Temperature and velocity field}
\label{sec:43_TemperatureVelocityField}
\begin{figure} 
	\centering
	\includegraphics[width=0.99\linewidth,trim=0 0 0 0,clip]{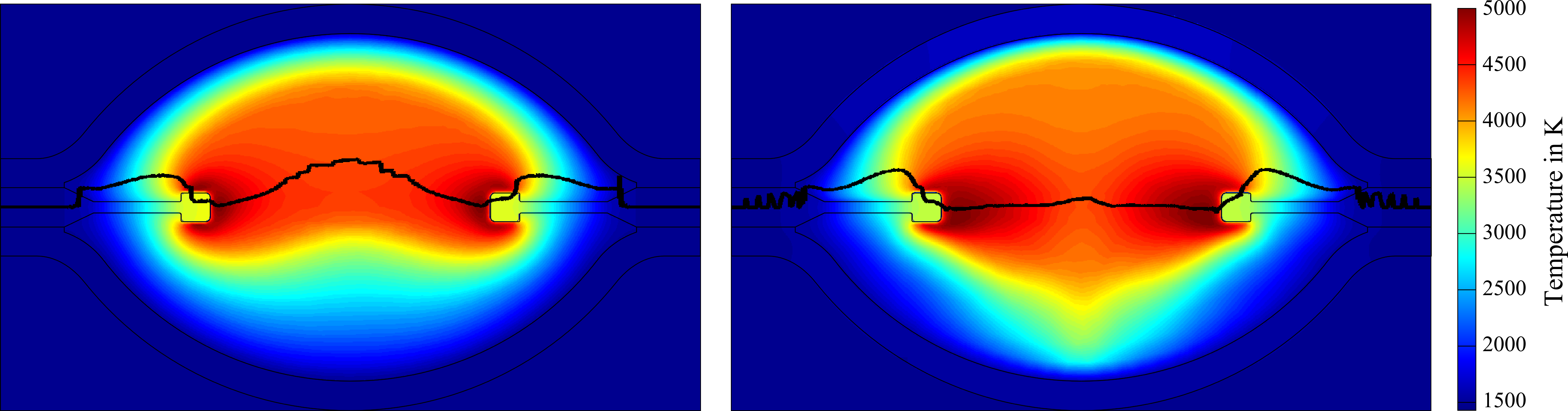}
	\caption{Temperature distribution in the $y$-$z$-plane at different excitation frequencies while decreasing the excitation frequency. Left: Far off resonance (\SI{47750}{Hz}). Right: Near resonance (\SI{46313}{Hz}). The black line indicates the location of maximal temperature as a function of the $y$-coordinate. \label{fig:TemperatureFields}}
\end{figure}

Figure~\ref{fig:TemperatureFields} right shows the temperature field resulting from the simulation at an excitation frequency of \SI{46313}{Hz}, the frequency, where the voltage jump in the blue curve of figure~\ref{fig:VoltvsXF} appears (decreasing frequency). A major impact of AS on the temperature field can be observed, when comparing it to the temperature field obtained at \SI{47750}{Hz} (figure~\ref{fig:TemperatureFields} left). This frequency is far off an acoustic resonance and the AS force accordingly small.

While approaching the resonance, the maximum plasma temperature, located at the hot-spots in front of the electrodes, only increases slightly. Simultaneously, the temperature in the lower part of the arc tube, especially in the vertical center plane of the arc tube, considerably increases. These higher temperatures are caused by a flow that is directed in negative $z$-direction. The flow in the upper part is directed into the positive $z$-direction and leads to a steeper temperature gradient near the upper wall.

\begin{figure} 
	\centering
	\includegraphics[width=0.99\linewidth,trim=0 0 0 0,clip]{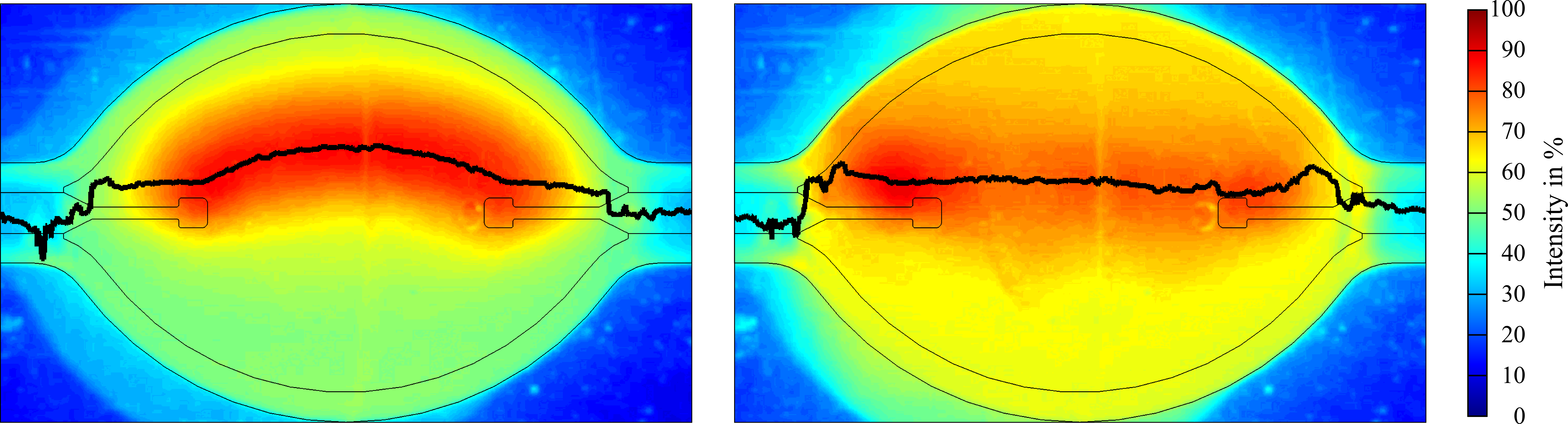}
	\caption{Measurements of the light intensity. The left part shows the plasma arc at stable operation conditions. The right part shows the plasma arc immediately before it starts to rotate violently around the lamp axis. The black line indicates the location of maximal intensity as a function of the $y$-coordinate. \label{fig:LightIntesity}}
\end{figure}

To compare the temperature field of the simulation with experimental results, the light intensity distribution was recorded by a camera \cite{Schwieger.2014}. The left part of figure~\ref{fig:LightIntesity} shows the intensity distribution at stable operation conditions. The buoyancy force bends the plasma arc upward and AS is negligible. The picture resembles the left part of figure~\ref{fig:TemperatureFields}, where the temperature distribution far off the resonance is depicted.

After tuning the parameters to unstable lamp operation, AS emerges. The intensity distribution on the right shows the discharge arc when it moves towards the lower part of the arc tube. Immediately afterwards, the discharge arc started to rotate around the $y$-axis and finally the lamp exploded. It highlights that AS acts against buoyancy. The pictures strongly resemble the right part of figure~\ref{fig:TemperatureFields} from the simulation.

\begin{figure} 
	\centering
	\includegraphics[width=0.99\linewidth,trim=0 0 0 0,clip]{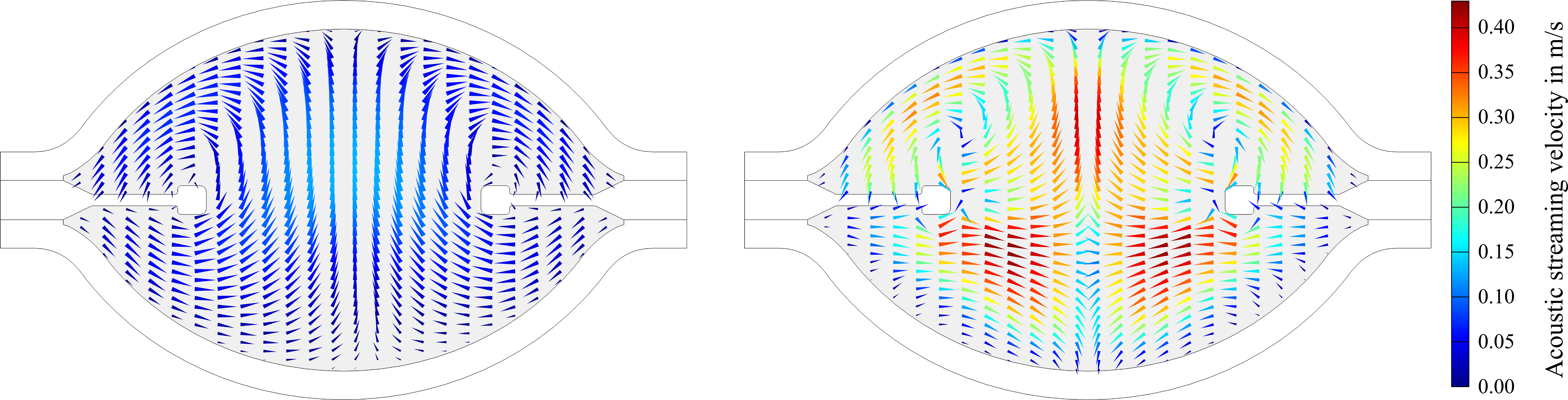}
	\caption{Fluid flow in the $y$-$z$-plane at different excitation frequencies while decreasing the excitation frequency: Left: Far off resonance (\SI{47750}{Hz}). Right: Near resonance (\SI{46313}{Hz}).}
	\label{fig:VelocityFields}
\end{figure}

Figure~\ref{fig:VelocityFields} shows the simulated flow fields corresponding to the temperature fields depicted in figure~\ref{fig:TemperatureFields}. At an excitation frequency far off the resonance the fluid flow essentially represents the purely buoyancy-driven flow field. The flow between the electrodes is directed upwards and has its maximum in this region. Instead of a maximal velocity of \SI{0.09}{m/s} in case of the purely buoyancy-driven flow, the maximal velocity is \SI{0.13}{m/s}. The flow field is not fully symmetric with respect to the $y$-$z$-plane due to the tilting of the lamp.

Near the resonance the AS dominates the buoyancy flow and the asymmetry caused by the lamp tilting is not visible any more because only the buoyancy force is affected by the tilting of the lamp. In the upper part of the arc tube two additional vortices develop (right: clockwise; left: counterclockwise). In the lower part the flow points in the direction of the negative $z$-axis. The flow continues in the $x$-direction, which is not visible in the figure. The maximal velocity is \SI{0.42}{m/s} whereas in the simulation without backcoupling it is \SI{0.61}{m/s} \cite{Baumann.2014b}. The reason for the difference is that in case of the coupled simulation the lamp is excited \SI{55}{Hz} off the resonance. The AS force is too weak to cause the symmetry breaking transition, which was observed in the uncoupled model \cite{Baumann.2014b}.

\section{Conclusions}
\label{sec:50_Conclusions}
In this work a stationary 3D finite element model is used to simulate the backcoupling of acoustic streaming on the temperature field inside a high-intensity discharge lamp. At the investigated acoustic eigenfrequency a hysteresis was detected in the simulation and confirmed by accompanying experimental investigations. The voltage drop when decreasing the excitation frequency shows a different behavior than in case of an increasing excitation frequency.
The simulation results show a severe impact of the acoustic streaming on the voltage drop, the temperature field and the flow field.
In the simulation as well as in the experiments the voltage drop at the investigated resonance significantly increases compared to a voltage drop far off the acoustic resonance.

AS causes the maximal velocity in the arc tube to become more than \num{4} times higher compared to the purely buoyancy-driven flow. A buoyancy dominated flow leads to an upward bending of the discharge arc. Acoustic streaming on the contrary induces a straight arc between the electrodes. These results are consistent with measurements of the light intensity distribution.
The results will help to understand the physical processes, that lead to flickering of the discharge arc inside high-intensity discharge lamps. They enable development of new lamp systems with improved energy-efficiency.

\vspace{3mm}
{\noindent\bf Acknowledgment:} 
This research was supported by the German Federal Ministry of Education and Research (BMBF) under project reference 03FH025PX2 and Philips Lighting. We are indebted to our colleague Klaus Spohr for discussions.

\bibliographystyle{unsrt}
\bibliography{HIDLiteratur}

\begin{thebibliography}{10}

\bibitem{Zissis.2010}
Georges Zissis and Marco Haverlag.
\newblock {D}iagnostics for electrical discharge light sources: {P}ushing the
  limits.
\newblock {\em Journal of Physics D: Applied Physics}, 43(23):230301, 2010.

\bibitem{Stoffels.2006}
Winfried~W. Stoffels, A.~H. F.~M. Baede, Joost A.~M. van~der Mullen, Marco
  Haverlag, and Georges Zissis.
\newblock {D}efinition of a high intensity metal halide discharge reference
  lamp.
\newblock {\em Measurement Science and Technology}, 17(11):N67--N70, 2006.

\bibitem{Waide.2006}
Paul Waide and Satoshi Tanishima.
\newblock {\em {L}ight's labour's lost: {P}olicies for energy-efficient
  lighting}.
\newblock OECD and International Energy Agency, Paris, 2006.

\bibitem{NEMA.2010}
{National Electrical Manufacturers Association}.
\newblock {N}{E}{M}{A} standards publication {L}{S}{D} 54-2010: {T}he strengths
  and potentials of metal halide lighting systems.
\newblock Rosslyn, 2010.

\bibitem{Trestman.2002}
Grigoriy~A. Trestman.
\newblock {M}inimizing cost of {H}{I}{D} lamp electronic ballast.
\newblock In IEEE, editor, {\em {P}roceedings of the {A}nnual {C}onference of
  the {I}ndustrial {E}lectronics {S}ociety}, pages 1214--1218, Piscataway,
  2002. IEEE.

\bibitem{Schwieger.2014b}
J{\"o}rg Schwieger, Marcus Wolff, Bernd Baumann, Freddy Manders, and Jos
  Suijker.
\newblock {C}haracterization of {D}ischarge {A}rc {F}licker in
  {H}igh-{I}ntensity {D}ischarge {L}amps.
\newblock {\em IEEE Transactions on Industry Applications}, in press.

\bibitem{Afshar.2008}
Farhang Afshar.
\newblock {T}he theory of acoustic resonance and acoustic instability in
  {H}{I}{D} lamps.
\newblock {\em Journal of Illuminating Engineering Society of North America},
  5(1):27--38, 2008.

\bibitem{Baumann.2014b}
Bernd Baumann, J{\"o}rg Schwieger, Marcus Wolff, Freddy Manders, and Jos
  Suijker.
\newblock {N}umerical investigation of symmetry breaking and critical behavior
  of the acoustic streaming field in high-intensity discharge lamps.
\newblock {\em Journal of Physics D: Applied Physics}, in press.

\bibitem{Kreuzer.1977}
Lloyd~B. Kreuzer.
\newblock {T}he physics of signal generation and detection.
\newblock In Yoh-Han Pao, editor, {\em {O}ptoacoustic spectroscopy and
  detection}, pages 1--25. Academic Press, New York, 1977.

\bibitem{Baumann.2007}
Bernd Baumann, Marcus Wolff, Bernd Kost, and Hinrich Groninga.
\newblock {F}inite element calculation of photoacoustic signals.
\newblock {\em Applied Optics}, 46(7):1120--1125, 2007.

\bibitem{Baumann.2009}
Bernd Baumann, Marcus Wolff, John Hirsch, Piet Antonis, Sounil Bhosle, and
  Ricardo~V. Barrientos.
\newblock {F}inite element estimation of acoustical response functions in
  {H}{I}{D} lamps.
\newblock {\em Journal of Physics D: Applied Physics}, 42(22):225209, 2009.

\bibitem{Boluriaan.2003}
Said Boluriaan and Philip~J. Morris.
\newblock {A}coustic streaming: {F}rom {R}ayleigh to today.
\newblock {\em International Journal of Aeroacoustics}, 2(3):255--292, 2003.

\bibitem{Temkin.1981}
Samuel Temkin.
\newblock {\em {E}lements of acoustics}.
\newblock Wiley, New York, 1981.

\bibitem{Schuster.1940}
K.~Schuster and W.~Matz.
\newblock {\"U}ber station{\"a}re {S}tr{\"o}mungen im {K}undtschen {R}ohr.
\newblock {\em Akustische Zeitschrift}, 5:349--352, 1940.

\bibitem{Hamilton.2003b}
Mark~F. Hamilton, Yurii~A. Ilinskii, and Evgenia~A. Zabolotskaya.
\newblock {A}coustic streaming generated by standing waves in two-dimensional
  channels of arbitrary width.
\newblock {\em The Journal of the Acoustical Society of America},
  113(1):153--160, 2003.

\bibitem{Schwieger.2014}
J{\"o}rg Schwieger, Bernd Baumann, and Marcus Wolff.
\newblock {A}rc {S}hape of {H}igh-{I}ntensity {D}ischarge {L}amps: {S}imulation
  and {E}xperiments.
\newblock {\em International Journal of Thermophysics}, in press.

\bibitem{Cross.2009}
Michael Cross and Henry Greenside.
\newblock {\em {P}attern formation and dynamics in nonequilibrium systems}.
\newblock Cambridge University Press, Cambridge, 2009.

\bibitem{Thompson.1986}
J.~Michael~T. Thompson and Lawrence~N. Virgin.
\newblock Predicting a jump to resonance using transient maps and beats.
\newblock {\em {I}nternational {J}ournal of {N}on-{L}inear {M}echanics},
  21(3):205--216, 1986.

\bibitem{Schwieger.2013}
J{\"o}rg Schwieger, Bernd Baumann, Marcus Wolff, Freddy Manders, and Jos
  Suijker.
\newblock {I}nfluence of thermal conductivity and plasma pressure on
  temperature distribution and acoustical eigenfrequencies of high-intensity
  discharge lamps.
\newblock {\em {P}roceedings of the {C}omsol {U}sers {C}onference}, 2013.

\bibitem{Chhun.2011}
Labo Chhun, Pascal Maussion, Sounil Bhosle, and Georges Zissis.
\newblock {C}haracterization of acoustic resonance in a high pressure sodium
  lamp.
\newblock {\em IEEE Transactions on Industry Applications}, 47(2):1071--1076,
  2011.

\end{thebibliography}

\end{document}